\documentclass[aps,prl,superscriptaddress,twocolumn,tightenlines]{revtex4-1}
\usepackage{amssymb}
\usepackage{graphicx}
\usepackage{natbib}
\usepackage{epstopdf}
\usepackage{dcolumn}
\usepackage{bm}
\usepackage{makeidx}
\usepackage{enumerate}
\usepackage {graphicx}
\usepackage {amsmath}
\usepackage {amsfonts}
\usepackage {amssymb}
\usepackage {mathrsfs}
\newcommand{\ud}{\,\mathrm{d}}

\begin{document}
\title{Long range $p$-wave proximity effect into a disordered metal }
\author{Aydin Cem Keser}
\affiliation{Condensed Matter Theory Center, Department of Physics,
University of Maryland, College Park, MD 20742-4111, USA}

\author{Valentin Stanev}
\affiliation{Condensed Matter Theory Center, Department of Physics,
University of Maryland, College Park, MD 20742-4111, USA}

\author{Victor Galitski}
\affiliation{Joint Quantum Institute and Condensed Matter Theory Center, Department of Physics,
University of Maryland, College Park, MD 20742-4111, USA}
\affiliation{School of Physics, Monash University, Melbourne, Victoria 3800, Australia}
\date{\today }

\begin{abstract}
We use quasiclassical methods of superconductivity to study  the superconducting proximity effect from a topological $p$-wave superconductor into a disordered one-dimensional metallic wire.  We demonstrate that the corresponding Eilenberger 
equations with disorder reduce to a closed non-linear equation for the superconducting component of the matrix Green's function. Remarkably, this equation is formally equivalent to a classical mechanical system (i.e., Newton's equations), with the Green function corresponding to a coordinate of a fictitious particle and the coordinate along the wire corresponding to time. This mapping allows to obtain exact solutions 
in the disordered nanowire in terms of elliptic functions. A surprising result that comes out of this solution is that the $p$-wave superconductivity proximity-induced into the  disordered metal remains long-range, decaying as slowly as the conventional $s$-wave superconductivity. It is also shown that impurity scattering leads to the appearance of a  zero-energy peak.

\end{abstract}
\maketitle
{\it Introduction.} -- Superconducting heterostructures have attracted a lot of attention recently as possible hosts of Majorana fermions~\cite{Fu,JDSau,Lutchyn, Alicea,Oreg,Wimmer,Mourik, Deng, Das}. One of the important outstanding questions in the studies of these heterostructures is the interplay between topological superconductivity and disorder~\cite{Brouwer,Akhmerov,Wade,Alexandro}. Here we explore this issue focusing on the leakage of $p$-wave superconductivity into a disordered metal.  Na{\"\i}vely, it may not appear to be  a particularly meaningful question, because unconventional superconductivity is known to be suppressed by disorder per Anderson's theorem~\cite{Anderson}. However, Anderson's theorem is only relevant to an intrinsic superconductor and has little to do with a leakage of superconductivity. 

The linearized Usadel equations are standard tools in studies of proximity effects~\cite{Usadel,EfetovRw}. Their derivation, however, assumes 
 that an anisotropic component of the superconducting condensate's wave-function is small compared to the isotropic one, which is not the case in the systems we are interested in. Here, we focus on the more general Eilenberger equations~\cite{Eilenberger, QGF}, which allow us to straightforwardly model systems with complicated geometries, and varying degree of disorder. (In the context of topological superconductivity, similar approach has been used in Refs.~\cite{Neven, Abay, VS1, Hoi}.) We obtain exact solutions of these equations, and study superconducting correlations induced by proximity in a metallic wire. In particular, we demonstrate that the $p$-wave correlations can be surprisingly long-ranged, even in the presence of disorder. We also show that impurity scattering produces a zero-energy peak in the density of states (DOS).       




{\it Solution for $s$-wave and $p$-wave order parameters.}-- We study the quasiclassical Green's function $\hat g$, which is a matrix in both Nambu and spin space~\cite{QGF}. It is obtained from the full microscopic Green's function by integrating over the energies close to the Fermi surface, and it faithfully captures the long lengthscale  features of the system~\cite{Maslov}. In one-dimensional systems, $\hat g$ depends on the Matsubara frequency ($\omega$), the center-of-mass coordinate of the pair $(x)$, and the direction of the momentum at the Fermi points ($\zeta={\bf p}_x/p_F=+1/-1$ for right/left going particles). The Green's function obeys the Eilenberger equation~\cite{Eilenberger,QGF,EfetovRw} 
\begin {equation}
\zeta v_F \partial_x \hat g = - [\omega \tau_3 , \hat g] + i[\hat \Delta, \hat g] - \frac{1}{2 \tau_{imp}} [\langle \hat g \rangle,\hat g]. 
\label{eilen}
\end {equation}
 The effect of impurities enters the equation through the mean time between collisions $\tau_{imp}$, and $\langle ... \rangle$ denotes an average over the Fermi surface. We ignore self-consistency, and assume that the order parameter $\hat \Delta$ is constant throughout the wire. (We believe enforcing self-consistency would not change our results qualitatively.)
 
We consider $s$-wave and $p$-wave order parameters in parallel, even though the appropriate Eilenberger equations differ significantly. First, we decompose the Green's function in Nambu space using the Pauli matrices $\tau_i$: $\hat g = -i g_1 \hat \tau_1 +  g_2 \hat \tau_2 + g_3 \hat \tau_3$. The scalar functions $g_i$ have to satisfy the normalization condition $-g_1^2 + g_2^2 + g_3^2=1$ (This will be referred to as the norm of $\hat{g}$, from here on). Note that the DOS of the system can be obtained from the diagonal  component $g_3$~\cite{VS1}.

In the case of an $s$-wave superconductor, $\hat \Delta$ is a spin-singlet and, ignoring the spin indices, it can be written as $\Delta_0 i \tau_2$. The diagonal component $g_3$ contains the particle-hole correlations.  The function $g_2$  encodes the $s$-wave pairing, whereas $g_1$ describes the $p$-wave, odd-frequency superconducting correlations, induced by boundaries or other inhomogeneities (it disappears in the bulk uniform state~\cite{Efetov,Buzdin,Golubov}).
In the case of a $p$-wave wire we consider spinless fermions, and the order parameter can be written as $\zeta \Delta_0 i \tau_2$. The difference from the $s$-wave case arises from the fact that now $g_2$ is $p$-wave, and $g_1$ contains the secondary $s$-wave (odd-frequency) correlations~\cite{Golubov, TanakaR, OFP}. 

The components of $\hat{g}$ obey three coupled differential equations. These equations, however, differ for the $s$-wave and the $p$-wave cases, due to the Fermi surface averaging: in the $s$-wave case we have $\langle g_1 \rangle=0$, $\langle g_2 \rangle=g_2$, whereas in the $p$-wave case $\langle g_1 \rangle=  g_1$, $\langle g_2 \rangle=0$. In both cases $\langle g_3 \rangle=g_3$ applies (particle-hole correlations are $s$-wave-like). We use an index $j = (1,2) $ that allows us to write the component equations in a unified way; in the $s$-wave case we have $j=1$, and $j=2$ pertains to the $p$-wave case. 
This index will be used for the rest of the paper, unless the state is explicitly indicated with a subscript $s$ or $p$. For the order parameters we have $\Delta_{(1)} \equiv \Delta_s = \Delta_0$ for $s$-wave, and $\Delta_{(2)} \equiv \Delta_p  = \zeta \Delta_0$ for $p$-wave. In $g_j$ the subscript denotes the Nambu space components -- $g_1$ and $g_2$ for $s$-wave and $p$-wave cases respectively. With these, and using the Kronecker delta $\delta_{i j}$, we write the Eilenberger equation as: 
\begin {subequations}
\label{first}
\begin {align}
\label{first_1}
&\zeta v_F \partial_x  g_1 = -2\omega g_2 +2 \Delta_{(j)} g_3 - \left( \frac{1}{\tau_{imp}} g_2 g_3\right)\delta_{j 2},\\
\label{first_2}
&\zeta v_F \partial_x g_2 = -2 \omega g_1 -\left(\frac{1}{\tau_{imp}} g_1 g_3\right)\delta_{j 1}, \\
\label{first_3}
&\zeta v_F \partial_x g_3 = 2 \Delta_{(j)} g_1 -(-1)^{j}\frac{1}{\tau_{imp}} g_1 g_2.
\end {align}
\label{system}
\end {subequations}  
In the clean case, these equations become linear and are easily solved~\cite{Efetov,Buzdin,VS1}. Impurities introduce nonlinear coupling, proportional to $1/\tau_{imp}$. Nevertheless, as we will demonstrate, these equations can still be treated analytically.

To be integrable, this system (either for $s$-wave or $p$-wave state)
should have two constants of integration. The norm of $\hat{g}$ is one of them , and it can be shown  that another constant is given by:
\begin {equation}
\label{constant_of_motion}
C_{(j)} = \frac{(-1)^{j-1}}{ 2\tau_{imp}}g_j^2 + 2 \Delta_{(j)} g_2 + 2\omega g_3.
\end {equation} 
This can be seen from equations Eq.~\eqref{first}, by verifying that $\partial_x C_{(j)}=0$, for  both  $s$- and $p$-wave cases. Using $C_{(j)}$ we can derive from the system (Eqs.~\ref{first}) a second-order equation for a  \emph{single} component. 
In the $s$-wave case we proceed by differentiating Eq.~\ref{first_1}. Using $C_s\equiv C_{(1)}$ we obtain the following equation:
\begin{equation}
v_F^2 \partial_{x}^2 g_1 =4 \alpha_s g_1 - \frac{g_1^3}{2\tau_{imp}^2},
\label{EOM_full_s}
\end{equation}
where we have defined $\alpha_{s} = \Omega^2 + C_s/(4\tau_{imp}) $, with $\Omega^2 = \omega^2 +\Delta_0^2$.
In the case of a $p$-wave order parameter we differentiate Eq.~\ref{first_2}, and by using $C_p\equiv C_{(2)}$, and defining $\alpha_{p} = \Omega^2 + C_p/(4\tau_{imp}) $, the resulting equation is:
\begin{equation}
v_F^2 \partial_{x}^2 g_2 =-2 \zeta \Delta_0 C_p + 4\alpha_p g_2 -\frac{3\zeta \Delta_0}{\tau_{imp}} g_2^2 +\frac{g_2^3}{2\tau_{imp}^2}.
\label{EOM_full_p}
\end{equation}
Either of these equations can now be integrated on its own, without {\it explicit} reference to the other two components. However, once $g_j$ is determined the other components follow from $C_{(j)}$ and $\partial_x g_j$.

We also consider the case of a normal metallic segment in contact with a superconductor with order parameter $\Delta_0$ (for $s$-wave) or $\zeta \Delta_0$ (for $p$-wave). 
To study the superconducting correlations induced in the normal part we can use the
Eilenberger equation with the order parameter in the metal set to zero. The constant of integration 
becomes $C_{(j)} = (-1)^{j-1} g_j^2/( 2\tau_{imp}) + 2\omega g_3$.
To streamline notation we introduce the dimensionless constants $\tilde{C}_{(j)} = 
C_{(j)}/2\Delta_0$, $\tilde{\alpha}_{(j)} = 
\alpha_{(j)}/\Delta_0^2$, and $\beta = 1/(2\tau_{imp}\Delta_0)$. In addition, we 
define the coherence length, $\xi_0 = v_F/\Delta_0$. (Note that in these definitions 
$\Delta_0$ is introduced only as an energy scale.) With these, we can write, for the 
$g_1$ component in a normal segment in contact with $s$-wave wire, the following 
equation: 
\begin{equation}
\xi_0^2 \partial_{x}^2 g_1 =4 \tilde{\alpha}_s g_1 -2\beta^2 g_1^3.
\label{EOM_s}
\end{equation}
In the case of a normal wire in contact with a $p$-wave superconductor we have equation for $g_2$:
\begin{equation}
\xi_0^2 \partial_{x}^2 g_2 =4 \tilde{\alpha}_p g_2 +2\beta^2 g_2^3.
\label{EOM_p}
\end{equation}
Notice the difference in the sign between the $\beta^2$ terms in the two equations. 
\begin{figure}[h]
\begin{center}$
\begin{array}{cc}
\includegraphics[width=0.45\textwidth]{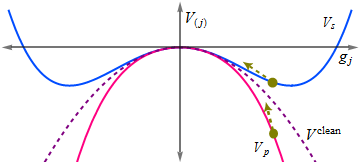}
\end{array}$
\end{center}
\caption{The potential landscape of a classical particle with motion describing the Green's function, for a normal metallic segment, in contact with a superconductor. Depending on the superconductor ($s$ or $p$-wave) potential is either $V_s$ or $V_p$. In the clean limit both converge to $V^{clean}$.}
\label{Fig1}
\end{figure}

{\it Classical particle analogy}.--  Equations  \ref{EOM_full_s}, \ref{EOM_full_p}, \ref{EOM_s} or \ref{EOM_p} can be integrated analytically. 
Before we do this, however, it is instructive to interpret them as equations of motion for a classical particle with one degree of freedom, moving in an external potential. The ``position'' of this particle is $g_{j}$  and the ``time''  $\tilde{t}$, is given by $ \zeta 2 x/\xi_0$, hence its ``momentum'' is $\partial_{\tilde{t}} g_j$. 
In both $s$-wave and $p$-wave cases the external potential is described by a quartic polynomial function. For example, from Eqs.~\ref{EOM_s} and \ref{EOM_p} we can write $V_{(j)}(g_j) = - \tilde{\alpha}_{(j)} g_j^2/2 + (-1)^{j-1} \beta^2 g_j^4/8$. Note that $V_j$ describes a double well for the $s$-wave case ($j=1$), and a
hill for the $p$-wave case ($j=2$). In the clean limit we have $\beta \to 0$ and the potential energy becomes an inverted 
parabola, $V_{(j)}(g_j)= -\tilde{\alpha}_{(j)} g_j^2/2$, for both the $s$-wave and the $p$-wave cases (see Fig. \ref{Fig1}).

We denote the dimensionless ``energy" of the classical system by $\tilde{E}_{(j)}$. It is a constant of integration, and can be determined from the boundary conditions for $g_j$. 

Since we want to study proximity effects, we concentrate on Eqs.~\ref{EOM_s} and \ref{EOM_p}. After multiplying both sides with $v_F \partial_{x} g_j$ we integrate the equations two times. The result is the following elliptic integral, where the variable $x$ spans the length of the wire that starts at $\ x=0$ and ends at $L$
\begin {equation}
\int\limits_{g_j(0)}^{g_j( x)}  \ud g_j\left( \tilde{\alpha}_{(j)} g_j^2 +(-1)^{j}  \frac{\beta^2}{ 4} g_j^4 + 2 \tilde{E}_{(j)}  \right)^{-1/2} = \pm \frac{2 x}{\zeta \xi_0}.
\label{integral1} 
\end {equation}
 The $\pm$ sign before the right hand side of Eq.~\ref{integral1} is to ensure that $x$ is positive, and it  depends on the choice of the integration contour in the complex $g_j$  plane.  We will denote the poles of the integrand as $\rho^\pm_{(j)}$. The integral can be written in terms of the inverse Jacobi elliptic function $\text{sn}^{-1}$, with elliptic parameter $m=\rho^+_{(j)}/\rho^-_{(j)}$. The monotonic solution is given by
 \begin {subequations}
 \label{jacobi_elliptic_sp}
 \begin {align}
&\text{sn}^{-1}\left( \frac{ g_j( x')}{(\rho^+_j)^{1/2}}\bigg\vert \frac{\rho_{(j)}^+}{\rho_{(j)}^-}\right)\bigg\vert^{x}_0  =  \pm\frac{\zeta \beta x}{\xi_0}[(-1)^{j}\rho^-_{(j)}]^{1/2}\label{sns},\\
 &\rho^\pm_{(j)} = \frac{2}{\beta^2} \left((-1)^{j-1}\alpha_{(j)} \pm \left[\alpha_{j}^2 -(-1)^{j} 2\tilde{E}_{(j)} \beta^2\right]^{1/2}\right).
 \end {align}
 \end {subequations}
It is important to note that another choice of the integration contour may lead to non monotonic, and/or oscillatory 
solutions. 
We can understand this by considering the classical 
particle in one of the potentials shown on Fig~\ref{Fig1}. In the $s$-wave case, the 
potential is a double well, hence the motion is generally periodic. However, the non-monotonic solutions are unphysical and we have to discard them, since the turning points of the trajectories scale as
$\pm(\omega \tau_{imp})$ at high frequency, and for both $\omega \to \infty$ or $\tau_{imp} \to \infty$ the periodic motion has unbounded amplitude. In the $p$-wave case, the period of the elliptic function is 
imaginary, as $V_p$ does not 
lead to periodic motion. We conclude that in both of the $s$-wave and $p$-wave cases the only physically acceptable solutions are monotonic (given by Eq.~\ref{jacobi_elliptic_sp}).
They can be visualized by imagining the motion of a particle, with initial position $g_j(0)$ and velocity directed towards the origin $g_j=0$,
climbing a non-harmonic hill potential $V_{(j)}(g_j)$. The amount of ``time'', for the particle to reach its final position represents the length of the wire $L$. For example, if $L$ is infinite the particle is coming to a stop at the origin (no superconducting correlations at infinity means vanishing velocity), hence should have zero ``energy'', $\tilde{E}=0$.

{\it $p$-wave wire with normal segment.}--  Let us use the solution of the Eilenberger equation to study the leakage of superconductivity in a metallic wire.  We consider an infinite wire extending along the $x$-axis with two segments that meet at $x=0$. 
The semi-infinite segment on the left ($x<0$) is made of clean $p$-wave superconductor. 
The segment on the right ($x>0$) is made of a diffusive normal metal (the order parameter is zero).
 
We obviously want a solution that, in the limit $x \rightarrow - \infty$ reproduces the mean field result for a uniform clean $p$-wave superconductor. Introducing the parameter $B$ and the dimensionless variables $\tilde{\Omega} = \Omega/\Delta_0 $, $\tilde{\omega}=\omega/\Delta_0$, we can write such a solution~~\cite{Matsumoto1, Matsumoto2, VS1}: 
\begin {subequations}
\label{initial_p}
\begin {align}
g_1(x)&=(1/\tilde{\omega})[1-\tilde{\Omega}B]\exp(2\tilde{\Omega} x/\xi_0) , \\
g_2(x)&= \zeta (1/\tilde{\Omega}) \left(1-[1- \tilde{\Omega}B]\exp(2\tilde{\Omega}x/\xi_0)\right) ,\\
g_3(x)&=  \left\lbrace [1 -  \tilde{\Omega} B]/(\tilde{\Omega} \tilde{\omega})\right\rbrace
\exp(2\tilde{\Omega} x/\xi_0) +\tilde{\omega}/\tilde{\Omega}. 
\end {align}
\end {subequations}
$B$ has to be determined from the boundary conditions at $x=0$.
For simplicity, we will consider the case of perfectly transparent boundary there, which guarantees the continuity of the Green's functions~~\cite{Zaitsev}. 

Now we consider the  diffuse normal segment with infinite length. Then, for $x \rightarrow \infty$ we have $g_1 \rightarrow 0$, $g_2 
\rightarrow 0$ and $g_3 \rightarrow \text{sgn}(\omega)$. The constant of integration 
is $\tilde{C}_p =  [-\beta g_2^2/2 + \tilde{\omega} g_3] $, when normalized to $2\Delta_0$. Using the fact that $\tilde{C}_p(0) = \tilde{C}_p(x\to \infty) = |\tilde{\omega}|$, we immediately 
obtain $B =(1/\beta)[-1 +(1 + 2  \beta[\tilde{\Omega}  - |\tilde{\omega}|])^{1/2}] $, with $\tilde{\alpha}_p = \tilde{\omega}^2 + \beta|
\tilde{\omega}|$.


We can understand intuitively the behavior of $g_2$ by again invoking the classical analogy. The particle in potential $V_p$, with ``position'' $g_2$ where time is $\tilde{t} = 2 x/\xi_0$,
starts at $g_2(0)=\zeta B$, with velocity  $  \partial_{\tilde{t}} g_2(0) =-\tilde{\omega}\zeta g_1(0) =- \zeta (1 - \tilde{\Omega} B)$, and moves towards its unstable equilibrium point $g_1(+\infty)=0$, gradually slowing down until $ \partial_{\tilde{t}} g_2(+\infty)=0$. Thus, the trajectory of $g_2$ satisfies $\tilde{E}_{p}=0$.  The integral in Eq. \ref{integral1} is now straightforward, and defining the dimensionless constant $\kappa=[1+\beta^2 B^2/(4\tilde{\alpha}_p)]^{1/2} $, we can write the solution for $g_2$:
\begin {equation}
\label{infinite_length_soln}
g_2(x) = \frac{\zeta B}{\cosh(x/\xi') + \kappa \sinh(x/\xi')}.  
\end {equation}
Here $\xi' = \xi_0/( 2\tilde{\alpha}_p^{1/2})$ gives the effective decay length of the solution (at $T=0$). In physical units it is 
\begin{equation}
\xi'=\frac{v_F}{\sqrt{4\omega^2 + 2|\Delta_0 \omega/( \tau_{imp} ) |}}. 
\label{xi_infinite_length}
\end{equation}
In the dirty limit we have $\xi' = \sqrt{D/| \omega|}$, where $D$ is the diffusion coefficient. Finally, in the clean limit $g_2$ converges to $\zeta B \exp(-2|\tilde{\omega}|x/\xi_0)$, as expected~~\cite{VS1}.

The other two components of the Green's function can be derived from $g_2$ using $\tilde{C}_p$ and the Eilenberger equations: $g_1= -\zeta \xi_0 \partial_x g_2 /(2 \tilde{\omega})$ and $g_3 = \text{sgn}(\tilde{\omega}) + \beta g_2^2/(2 \tilde{\omega})$. As expected,  impurities suppress $g_2$ relative to $g_1$. However, they both decay in the normal segment over the \emph{same} lengthscale, given by Eq. \ref{xi_infinite_length}. This decay is long-range, and furthermore, with exactly the same lengthscale we obtain for the case of $s$-wave order parameter (see below). Thus, the na{\"\i}ve expectation of strong suppression of the $p$-wave correlations is misleading in this case. This is one of the main points of our paper. 
\begin{figure}[h]
\begin{center}$
\begin{array}{cc}
\includegraphics[width=0.45\textwidth]{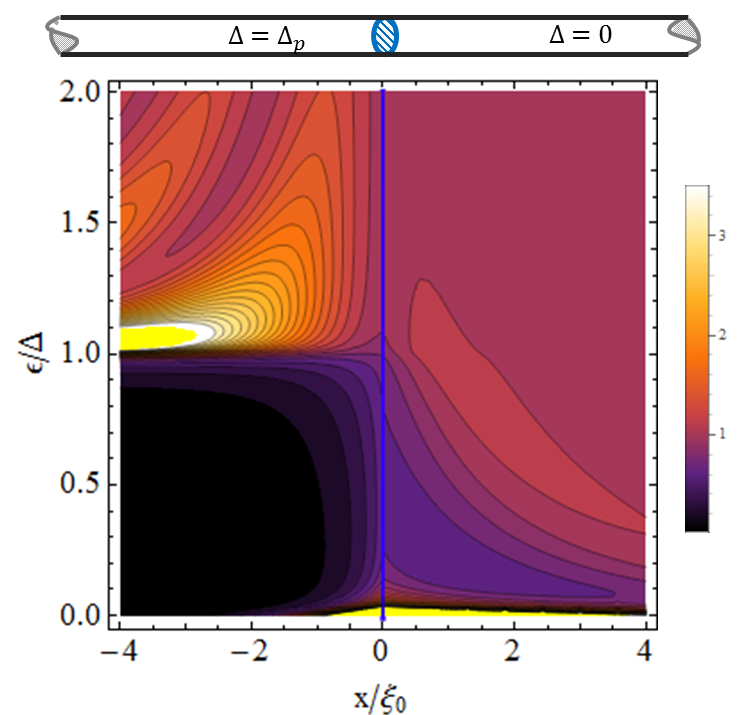}
\end{array}$
\end{center}
\caption{Contour plot of the DOS of an infinite wire. There is moderate disorder ($\beta=1$) in the normal segment ($x>0$). The solid yellow marks the regions that are beyond the plot range (where $N/N_0>3.5$). Notice the zero-energy peak in the normal segment.}
\label{Fig2}
\end{figure}

We can now obtain the DOS of the system, which is proportional to the real part of  $g_3(\omega\rightarrow -i\epsilon + \delta)$. On Fig. \ref{Fig2} we show the DOS for a system with moderate amount of disorder. Several things are apparent from this plot. First, for energies below $\Delta_0$ there is a significant decrease in the DOS of the normal segment, caused by the proximity effect; however, it is not a real gap, since the DOS stays finite everywhere. This decrease is entirely due to impurities -- in the clean case the DOS is constant for $x>0$~~\cite{VS1}. The impurity-induced term in $g_3$ also has a divergence in the limit of small frequencies ($g_3 \sim 1/\omega$), which leads to an infinite peak in the DOS. This zero-energy peak has the same origin as the Majorana edge state (namely, the sign change in the order parameter~\cite{Matsumoto1,Matsumoto2,ZBP}). Thus, in the infinite wire case, impurity scattering creates zero-energy peak, but it is not sufficient to localize it exponentially.  

As a side note, if the $p$-wave superconductor was replaced by an $s$-wave superconductor, the solution to Eq.\eqref{EOM_s} would be $g_1 =\zeta A [\cosh(x/\xi') + \kappa_s \sinh(x/\xi')]^{-1}$.  Here, $\kappa_s=[1-\beta^2 A^2/(4\tilde{\alpha}_s)]^{1/2} $ and $\zeta A$  is the value of $g_1$ at the junction, and is determined by the boundary values at the infinities in a way similar to that in the $p$-wave case. However, unlike the $p$-wave case, the $g_1 $ component at the boundary  is proportional to $\tilde{\omega}$. This dependence on $\tilde{\omega}$ changes the zero energy behavior of the DOS as follows. From $g_3 = \text{sgn}(\tilde{\omega})-\beta/(2 \tilde{\omega}) g_1^2$, we see that the low frequency limit is finite and thus there is no zero energy peak in the $s$-wave case~\cite{swave}.

If the normal segment has finite length $L$, we impose the condition $g_2(L) =0$, since the $p$-wave component is suppressed by the reflection from the boundary. Then the solution follows immediately from Eq.~\ref{jacobi_elliptic_sp}  as  $g_2(x) = \zeta (\rho_p^+)^{1/2}\text{sn}[\beta 
(\rho_p^-)^{1/2} (x-L)/\xi_0] $, with elliptic parameter 
$m=\rho_p^+/\rho_p^-$. However, this expression has limited practical value. The unknown constant $B_L$, which should be obtained from matching the two solutions for $g_2$ at $x=0$, enters the expression through the parameters $\rho_p^{\pm}$ , which makes it difficult to solve.
Fortunately, an approximate analytic form for $B_L$ can be obtained. In the limit $L 
\to \infty$, $B_L$ converges to $B$, that was previously calculated for the infinite wire case. In the opposite limit, $L \to 0$, $B_L$ vanishes. Numerical investigation suggests that $B_L$ as a function of $L$ can be approximated by $B [1 - \exp(-2 L/\lambda_B)]$, where the length scale $\lambda_B$ controls how quickly $B_L$ approaches to the infinite wire limit with increasing $L$. By expanding the integral in Eq.~\ref{integral1} around $B=0$ and matching it with the approximate expression,  we obtain $\lambda_B = B \xi_0$. 
\begin{figure}[h]
\begin{center}$
\begin{array}{cc}
\includegraphics[width=0.5\textwidth]{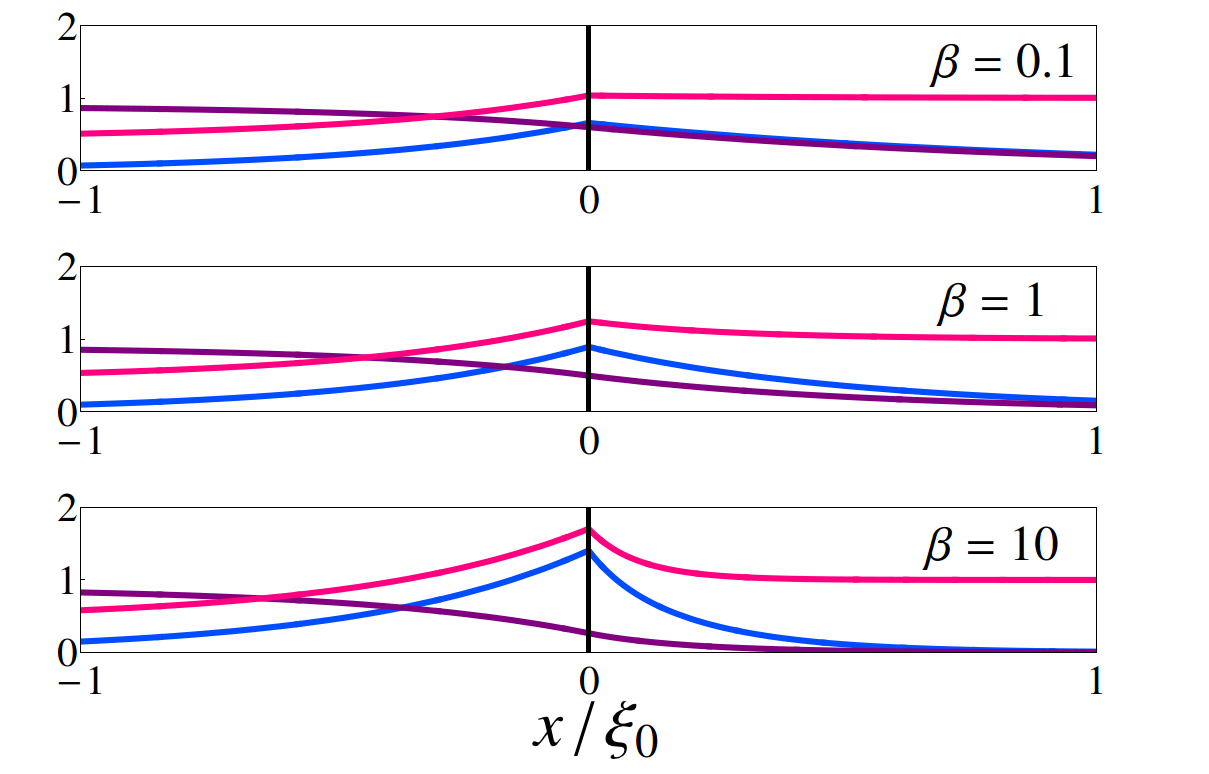}
\end{array}$
\end{center}
\caption{Components of $\hat{g}$,($g_1$: blue, $g_2$: purple, $g_3$: red) for a wire with infinite p-wave section and finite disordered section of length $L=5\xi_0$.
Top panel: weak disorder($\beta= 1/(2\tau_{imp}\Delta_0) = 0.1$). Middle panel: moderate disorder ($\beta=1$). And bottom panel: strong disorder ($\beta=10$). The Matsubara frequency is set to $\omega = \Delta_0/2$.}
\label{Fig3}
\end{figure}
Once we have $B_L$, we can use addition and transformation rules for elliptic functions~\cite{integral_table} to
write $g_2$ in a form that  manifestly converges  to that of the $L=\infty$ case. To save space, we shorten the common argument of elliptic functions,  $\beta |\rho_p^-|^{1/2} x/\xi_0$ as $(.)$. The common elliptic parameter of the elliptic functions is $(\rho_p^-  - \rho_p^+)/\rho_p^-$, and it lies in the interval $[0,1]$. With these definition we get:
\begin{equation}
g_2(x) = \zeta \frac{B_L \text{dn}(.) - \text{sn}(.) \text{cn}(.) \sqrt{|\rho_p^+| +B_L^2 } \sqrt{1+ B_L^2/|\rho_p^-|}}{\text{cn}^2(.) - (B_L^2/|\rho_p^-|) \text{sn}^2(.)}.
\label{finite_length_soln}
\end{equation}

We can again obtain the two other components from $g_2$ by using: $g_1= -\zeta \xi_0 \partial_x g_2 /(2 \tilde{\omega})$ and $g_3 = (\tilde{\alpha}_p - \tilde{\omega}^2)/(\beta \tilde{\omega}) + \beta g_2^2/(2 \tilde{\omega}) $. 

As $L\to\infty$, $\tilde{E}_p$ tends to zero, 
the elliptic functions are replaced by their  hyperbolic counterparts, and  we recover the solution for the infinite wire case (Eq.~\ref{infinite_length_soln}).

\begin{figure}[h]
\begin{center}$
\begin{array}{cc}
\includegraphics[width=0.4\textwidth]{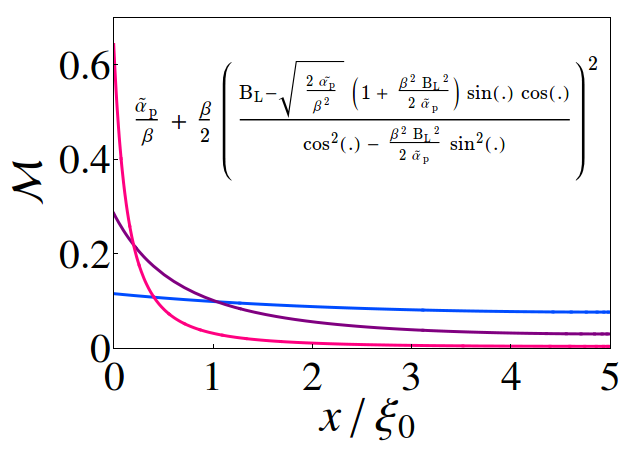}
\end{array}$
\end{center}
\caption{The weight of the zero energy mode $\mathcal{M}(x)$ in a normal section with length $L=5 \xi_0$ for three disorder strengths
 (blue:$\beta= 1/(2\tau_{imp}\Delta_0)=0.1$, purple: $\beta=1$, red: $\beta=10$). The expression in the inset is deduced from \eqref{finite_length_soln}, and $(.)$ stands for $(2\tilde{\alpha}_p
)^{1/2} x/\xi_0$. 
} 
\label{Fig4}
\end{figure}

Again, it is the impurity-induced contribution to $g_3$ that is of most interest. After analytic continuation we can write the zero-energy limit as:
\begin{equation}
g_3(x)=\frac{1}{\pi}\delta(\epsilon)\mathcal{M}(x).
\end{equation}
The function $\mathcal{M}(x)$ describes the $x$-dependent weight of the zero energy mode, and we can extract it from Eq.~\ref{finite_length_soln}. Its values at the junction point and at the end of the wire are $\mathcal{M}(0) = 1-B_L$ and $\mathcal{M}(L) = \mathcal{M}(0)-\beta B_L^2/2$ respectively. It can be approximated by a decaying exponent with decay length $\lambda_M =\xi_0 \beta B_L / (4 \tilde{\alpha}_p + 2\beta^2 B_L^2)$. Thus, in sharp contrast with the $L=\infty$ case, the zero-energy peak of a finite wire is exponentially localized. Figure~\ref{Fig4} shows $\mathcal{M}(x)$ in the normal section with length $L = 5 \xi_0$, for various disorder strengths. As can be seen, $\mathcal{M}(x)$ (i.e., the zero-energy peak) becomes more localized as the disorder  in the normal section increases.



This research was supported by DOE-BES DESC0001911 (VG \& VS), NSF-CAREER DMR-0847224 (ACK), and Simons Foundation.

 \bibliographystyle{apsrev}

\end{document}